# Ion beam shaping and downsizing of nanostructures

**M Zgirski[1], K-P Riikonen[1], V Tuboltsev[1,2], P Jalkanen[1], T T Hongisto[1] and K Yu Arutyunov[1]**

[1]NanoScience Center, Department of Physics, University of Jyväskylä, PB 35, FI-40014 Jyväskylä, Finland

[2]University of Helsinki, Department of Physical Sciences, Accelerator Laboratory, PB 43, Pietari Kalmin katu 2, FI-00014 Helsinki, Finland

E-mail: Konstantin.Arutyumov@phys.jyu.fi

**Abstract.** We report a new approach for progressive and well-controlled downsizing of nanostructures below the 10 nm scale. Low energetic ion beam ($Ar^+$) is used for gentle surface erosion, progressively shrinking the dimensions with ~ 1 nm accuracy. The method enables shaping of nanostructure geometry and polishing the surface. The process is clean room / high vacuum compatible being suitable for various applications. Apart from technological advantages, the method enables study of various size phenomena on the same sample between sessions of ion beam treatment.

PACS: 81.16.Rf, 37.20.+j, 62.23.Hj, 68.37.Ps, 34.50.Bw, 52.20.Hv, 34.50.D, 79.20.R

# 1. Introduction

Currently available portfolio of techniques which may in principle meet the requirement of delivering a surface feature size less than 100 nm is rather limited. Basically, these are lithographical techniques in which patterns are 'written' on a substrate with masked or finely focused optical (UV), X-ray, electron (EB) or focused ion beams (FIB). UV photolithography has been in active use for decades in semiconductor industry being well compatible with mass production [1]. Due to fundamental limitations, it is very unlikely that UV lithography can noticeably surpass sub-30 nm in the foreseen future [2-3]. State-of-the-art Electron Beam Lithography (EBL) was proved to be capable of delivering ~ 10 nm resolution. Unfortunately, EBL is slow, very expensive and it is very unlikely that it can effectively go below ~ 10 nm defined by the size of a molecule of an organic resist typically used for mask fabrication. The same limitations hold for X-ray and FIB with additional tremendous difficulties in developing equipment for beam manipulation and focusing on nm scales. An alternative is to use scanning nanosized probes, e.g. scanning probe or atomic force microscops (SPM and AFM), for patterning a resist or direct 'writing' (manipulation of nanoobjects). Though the approach may in principle provide resolution on a nm level, structures produced by such methods are rather unique and difficult to reproduce. Another type of lithography is based on the use of stamping or imprinting: NanoImprint Lithography (NIL) and Micro Contact Printing (μCP). There is an opinion [4] that the alternative lithography may provide a cost effective fabrication in the sub-50 nm range. However currently the stamping/impinting techniques are most likely to deliver reproducible resolution in a near sub-100 nm range.

Since it is hard to expect that in the nearest future existing techniques will be capable to deliver reliably sub-10 nm resolution, it is quite logical to consider an alternative: *further reduction of dimensions by post-processing of nanostructures obtained by conventional methods*. As far as sub-10 nm range is of concern, basically dry etching technique can be considered, as wet (chemical) processing is typically much less precise. Available dry etching techniques may be roughly subdivided into two major classes: Plasma Etching (PE) and Reactive Ion Etching (RIE). The working process is essentially the same: removal of surface atoms due to bombardment of species from plasma either by simple physical sputtering (PE) or chemically enhanced etching (RIE). Downsizing is achieved by gradual removal of material from the surface of a structure. Unfortunately, commercially available conventional PE and RIE apparatus are well known to suffer from the lack of precise control over processed structures with sub-100 nm dimensions. The core problem is unpredictable nature of local interaction between atomic species and surface features of 3D nanostructures when immersed into plasma. Ion beam sputtering can be also considered as plasma-based dry etching. It has been used for decades to reduce thickness of 2D objects (e.g. preparation of samples for TEM). To our knowledge, no research has been made to apply the method to modify the shape of 3D micro- or nanosystems.

In this paper we present a new approach based on low energetic wide ion beam etching to reduce dimensions of various types of micro- and nanoobjects in predictable and well-controlled way. The method is complementary to other nanofabrication processes. The technique can be used to obtain state-of-the-art small nanostructures or/and to study size phenomena on a single sample with progressively reduced dimension(s).

## 2. The method

Nanostructures were originally fabricated on SiO/Si substrates using conventional EBL on PMMA/MMA mask followed by a lift-off. After careful analysis with SPM the samples were subjected to irradiation by inert $Ar^+$ ions to course surface erosion by sputtering. The ions were produced in Tectra IonEtch ECR ion source, accelerated to 0.2 - 1 keV energy and delivered into a vacuum chamber with a base pressure of $10^{-7}$ mBar. While sputtering the pressure was kept about $10^{-4}$ mBar. The chamber was equipped with a multi-axis manipulator that allowed to tilt and to rotate the sample holder with respect to the beam incidence. The latter turned out to be essential in achieving a high degree of control in the etching processing. Total dose of ions impinging the sample was carefully monitored by ion beam current integration with resolution $\sim 10^{14}$ $1/cm^2$. Homogeneity of sputtering over the whole area of samples was ensured by in-beam rotation around the azimuth angle and beam wobbling. In case of high-resistive objects (e.g. mica) beam neutralizer was used preventing charging of the target.

Various types of systems have been successfully processed: insulators (sapphire, $Si/SiO_x$, mica), semicondctors (Si), metals (Al, Cu, Bi, Ti) [5-9] and hybrid metal-oxide-metal nanostructures (e.g. single electron transistors) [5]. In all cases the technique appeared to be non-destructive enabling study of size-dependent phenomena on the *same* sample between the sessions of the ion beam treatment. In this report we focus on technological aspects of the method applied to metallic nanowires. Starting from EBL-fabricated long strips with effective diameter $\sigma^{1/2} \sim 100$ nm ($\sigma$ being the wire cross-section) we were able to reduce the diameter down to $\sigma^{1/2} \sim 8$ nm. When approaching the 10 nm range, accuracy of the etching appears to be rather sensitive to the process parameters [10].

An example of evolution of dimensions (and shape) of a section of an aluminium nanowire after a sequence of sputtering is presented in figures 1 and 2. It is clear that the cross-section of the wire is gradually reduced. Co-sputtering of silicon substrate (figure 1(c)) suspending aluminium structure is more pronounced compared to metals with higher sputtering yield like bismuth, copper or tin. It should be mentioned that at low acceleration energies used in experiments (< 1 keV) the penetration depth of $Ar^+$ atoms into a metallic matrix is about few nm. For chemically active materials as aluminium and tin, the thickness of the radiation damage is comparable with the thickness of the naturally oxidized surface.

Comparison of the wire evolution being sputtered by $Ar^+$ ions impinging the target at the same angle (figure 1(a)) and from various directions (figure 1(b)), reveals an important feature. Fixed angle bombardment makes a structure less and less homogenous, while variation of the impingement angle 'polishes' the surface removing original geometrical imperfections. The effect is clearly visible on SEM images (figure 2): the granular structure typical for polycrystalline systems disappears after the ion beam treatment. We associate the effect with angular dependence of the sputtering rate applied to polycrystalline nanostructures. It is a known fact that sputtering yield (number of atoms ejected from the target per one incident ion) depends on the angle $\alpha$ between target's surface and projectiles' trajectory. Calculated dependence of the yield on angle for aluminium bombarded by 1 keV $Ar^+$ ions using SRIM software [11] is presented in figure 3 (open symbols). It should be noticed that density of projectiles at the sides of the wire is reduced by factor $\sim \cos(\alpha)$. The amount of the removed material is proportional to the density of projectiles. Hence, the yield (the actual rate of material removal) should be re-calculated accordingly (figure 3, solid symbols). It is clear that if to irradiate a 3D structure at a constant angle (inset in figure 3), then different points of the sample are sputtered with different rates. Since lift-off fabricated polycrystalline structures has been studied, it is not excluded that different grains of the sample can be sputtered with different rates being dependent on their crystallographic orientation with respect to the ion beam axis. An effective solution to get rid of the preferential sputtering is to use wobbling or rotating sample stage tilted with respect to the beam axis (figure 1(b)). Variation of the impingement angle eliminates differences in etching rate related to existence of the sample sides (shadow effect) and anisotropy of sputtering of grains forming a polycrystalline structure. In this configuration all points of a structure are sputtered homogeneously with an average rate shown in figure 3 with dashed horizontal line. This approach was used in our earlier studies of quantum size phenomena, where we were able to trace the evolution of electron transport properties on the same sample while progressive and uniform reduction of its characteristic dimension(s) [5-9].

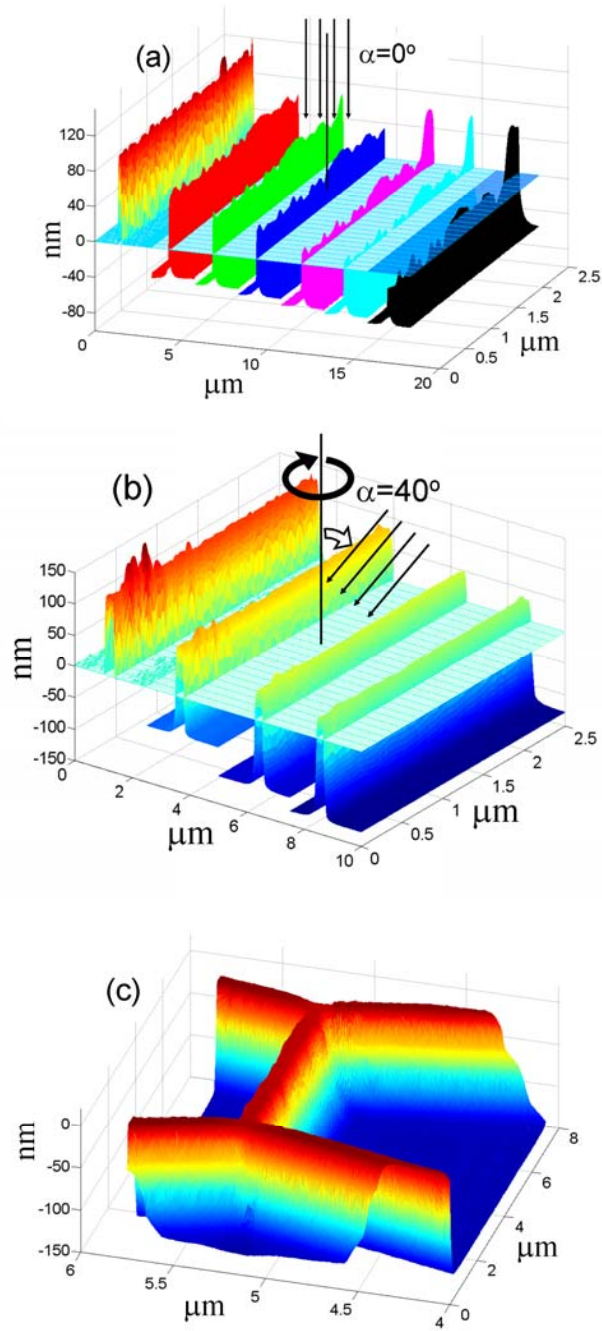

**Figure 1.** (a)-(b) SPM images showing evolution of the same aluminium nanowire after sessions of ion beam sputtering. Arrows indicate the direction of bombardment with 1keV Ar$^+$ ions. (a) Constant angle perpendicular to the plane of the structure. Note progressive development of structural inhomogeneity. (b) Sputtering of a similar nanowire at 40º while rotating the substrate. Note the 'polishing' effect removing the original sample roughness. Plane with grating (height = 0) separates silicon from aluminium. (c) SPM image of the *Si/SiOx* pedestal after several sessions of ion beam sputtering when the remaining metal has been removed with 7 % *HCl* water solution. The data was collected with *VEECO Dimension 3100* AFM in tapping mode using *NanoSensors* PPP-NCHR silicon cantilevers with tip radius <10 nm.

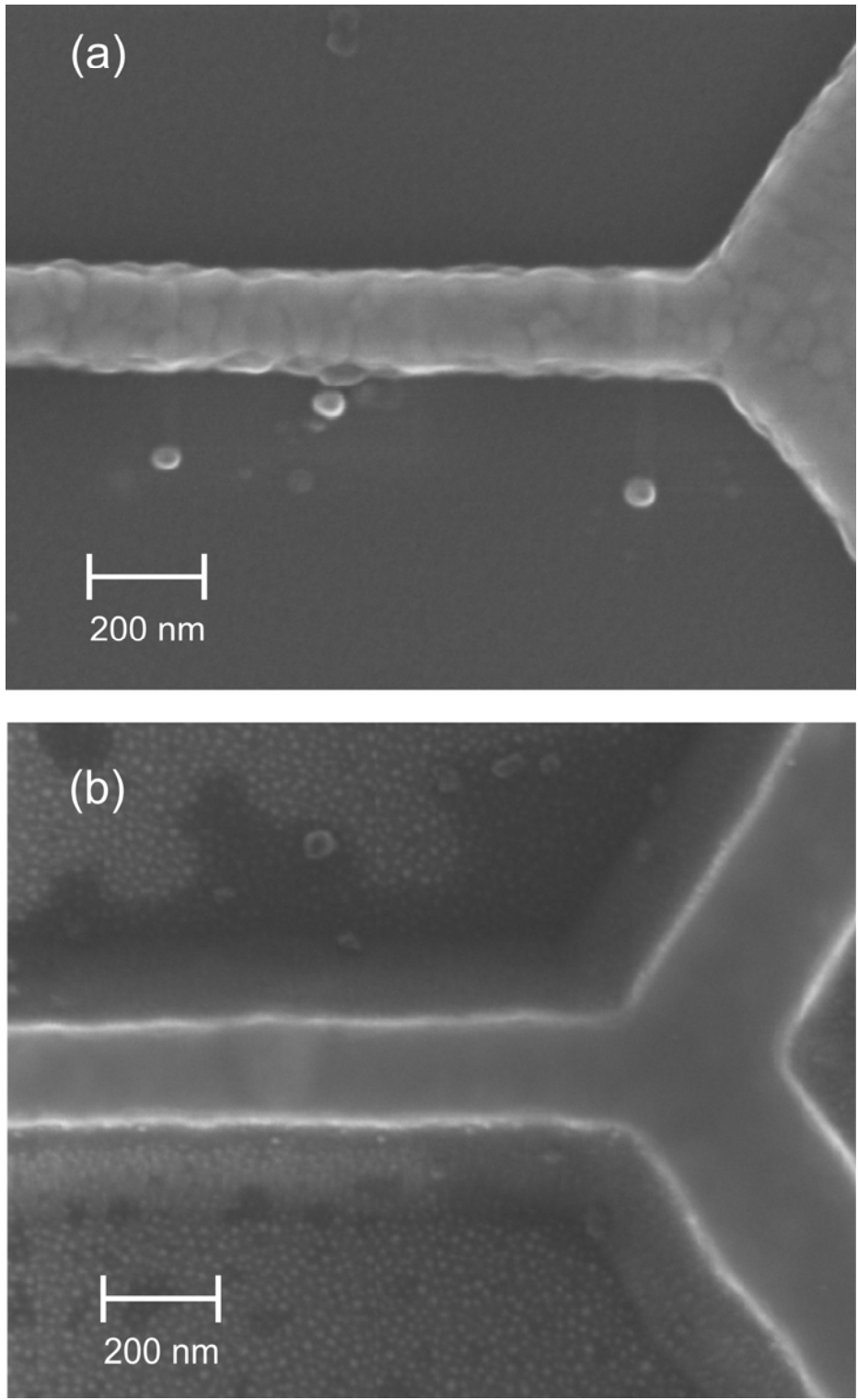

**Figure 2.** SPM images showing the same aluminium nanostructure before (a) and after (b) a session of Ar$^+$ ion bombardment at various impingement angles. One can clearly see the removal of the initial surface granularity typical for polycrystalline systems. Note the development of the wide 'shadow' along on the processed sample: sputtered *Si/SiO$_x$* substrate forming a pedestal supporting the metal on top. The images were taken with *e-Line* SEM using secondary electron emission at 10 keV e-beam acceleration energy.

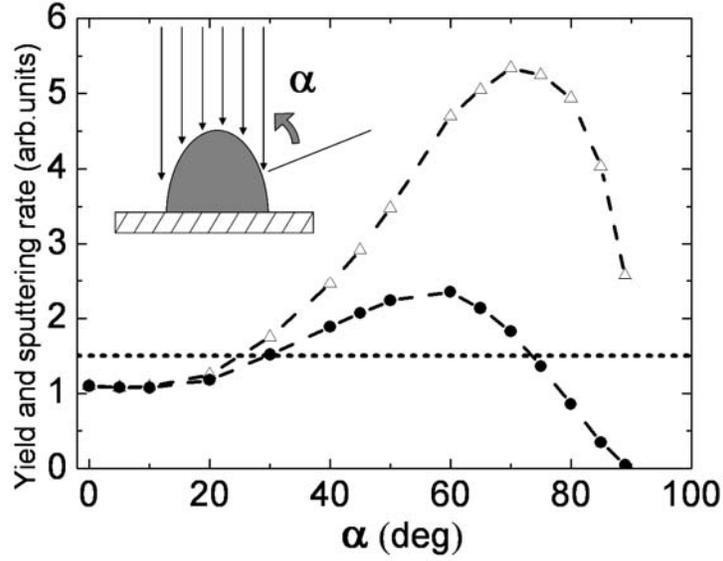

**Figure 3.** Angular dependence of yield and effective sputtering rate. Calculations are made for 1keV Ar$^+$ ions impinging aluminium target. Open symbols stand for the yield, solid ones - for the effective rate taking into account the reduction of the ion density at the sample's slopes by factor $\sim \cos(\alpha)$. The inset explains the geometry. Dashed horizontal line corresponds to the effective sputtering rate averaged over all possible angles.

The method enables not only the reduction of dimensions of pre-fabricated nanostructures, but also the modification of their original shape. Utilizing the angular dependence of the sputtering rate, one can model the evolution of the shape of a nanostructure after a particular dose of ion bombardment. Cross-section of lithographically-fabricated nanostructures (e.g. wires) typically has Gaussian shape with flattened top. For purposes of modelling with a good accuracy this shape can be approximated by a trapezoid (figure 4). To avoid development of microroughness, as discussed above, it is preferable to varynate the bombardment angle by rotating the tilted stage. In this case, the top (flat) part of a wire is exposed to the ion beam all the time being sputtered at a fixed rate $V_{top}$ dependent on the impingement angle. The sides are exposed only half time (being in a shadow for the second half period of the stage rotation) and are sputtered with the averaged rate $V_{side}$ (horizontal line in figure 3). One can show that depending on the initial ratio between the height $h$ and width $w$ of the trapezoid, the sputtered structure becomes more flat or contrary sharper (figure 4). The evolution is defined by the initial geometry: height $h$, width $w$ and the side wall angle $\theta$. If $(h/w)_{init} < \dfrac{\sin\theta}{2} \cdot \dfrac{V_{top}}{V_{side}}$ then flattening scenario takes place (figure 4(a)), while in the opposite limit the trapezoid becomes 'sharper' finally degenerating into a triangle (figure 4(b)). In both cases the dependence of the cross-section area $\sigma$ on total dose of impinged ions Φ has a parabolic dependence (figure 4). For typical structures used in our experiments the ratio $V_{side} / V_{top} \sim 0.4$. Hence, we always observed the flattening scenario. To obtain sharpening one should fabricate a sample with initial aspect ratio $h/w > 1.25$ and the side angle θ approaching 90°.

Though the proposed model of a nanostructure shape evolution is fairy simple, it gives reasonably good agreement with experiment. Dependence of the cross section of a real aluminium nanowire being bombarded by $Ar^+$ ions on rotating stage tilted at 40° is presented in figure 5. Experimental points are obtained by averaging the measured cross-sections while multiple SPM scans across the length of the wire. Relatively high error in determination of the wire cross-section originates from the uncertainty in definition of the boundary between the metal and the sputtered silicon substrate. Solid lines in figure 5 are fits to the abovementioned model with the only adjustable parameter for a given geometry being the sputtering rate at the top of the wire $V_{top}$. We performed similar experiments on silicon, tin, titanium and bismuth nanowires and various metal-oxide-metal hybrid nanostructures. The obtained results are qualitatively similar to aluminium used in this paper as a representative example.

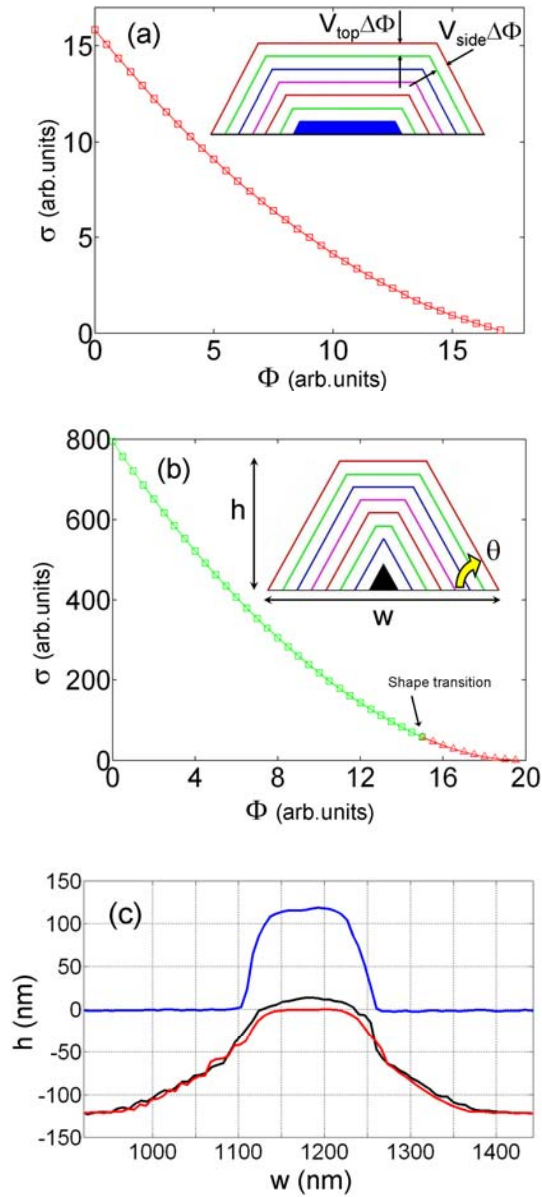

**Figure 4.** (a)-(b) Calculated variation of cross-section of an idealized nanostructure with trapezoid shape on ion sputtering dose $\Phi$. Depending on the initial sample geometry ($h$, $w$ and $\theta$) and the ratio of sputtering rates on top and on sides ($V_{top} / V_{side}$) for a given angle of incidence $\alpha$, there are two scenarios of the shape evolution: flattening (a) and sharpening (b). Here the simulation is made for the simplest case $V_{top} = V_{side}$ and $\theta = 60°$. Ion dose $\Phi$ is defined as the flux of projectiles ($Ar^+$ ions) integrated over the time of sputtering. (c) SPM measured typical cross-sections of the same structure: top curve – before sputtering, middle – after sputtering, bottom – $Si/SiO_x$ pedestal after removal of aluminum in 7% *HCl* water solution. Level "0" corresponds to the initial boundary between the $Si/SiO_x$ substrate and the metallic nanostructure. The SPM data was collected with *VEECO Dimension 3100* AFM in tapping mode using *NanoSensors* PPP-NCHR silicon cantilevers with tip radius <10 nm.

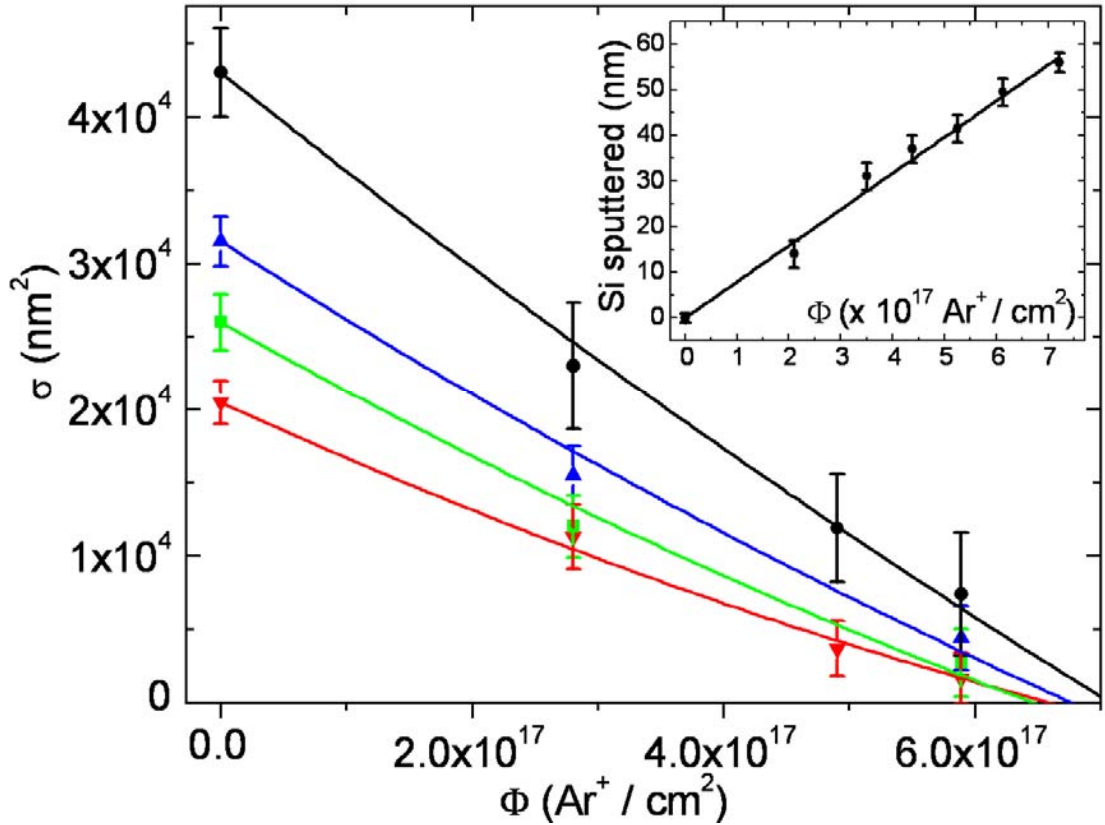

**Figure 5.** Evolution of cross-sections σ of several aluminium nanowires with sputtering ion dose Φ. Experimental points were obtained from SPM measurements. Solid lines are calculations based on the 'flattening scenario' model discussed in the text and in figure 4(a). Inset shows experimental data on sputtering of a flat Si substrate.

**3. Characterization of metallic samples**

The described method is applicable to a wide range of micro- and nanostructures whenever there is a necessity to re-shape the original object. When carefully calibrated, the ion etching provides reproducible accuracy ~ 1 nm. It is desirable to have an independent <u>non-destructive</u> characterisation enabling control of the structure dimensions after (or even - while) the ion beam processing. Unfortunately, at sub-10 nm scales this is not a trivial task. Scanning electron microscopy (SEM) often provides a poor contrast between a nanoobject made from a 'light' material (e.g. aluminium) and a substrate (figure 2). Extra difficulty comes if the substrate is highly insulating (e.g. mica) resulting in charging of the sample under the electron beam. Formally SPM characterisation should be immune to the mentioned drawbacks. The main problem with SPM analysis comes from inability to reliably distinguish a boundary between a 3D sample and a substrate. After several sessions of the ion beam sputtering both the sample <u>and</u> the substrate are sputtered resulting in a rather high aspect ratio system (figure 1(c)). Sensitivity of an SPM on the 'slope' of such a structure is much lower than for a flat sample. Removal of the material from the substrate (e.g. by chemical etching) can solve the problem (as in figure 1(c) and 4(c)), but the sample is destroyed.

In case of highly conducting (e.g. metallic) nanostructures electric resistance measurement can provide a useful and rather accurate characterisation of the sample dimensions. Determination of thickness of a 2D system (thin film) from its resistance is rather trivial. Of particular interest of the present paper was the determination of cross-section $\sigma$ of quasi-1D metallic nanowires by measuring their electric resistance $R=\rho L/\sigma$. The length of a sample L can be easily measured, for example, by SEM, while the cross-section $\sigma$ is the subject of interest. It is a textbook knowledge that for metals the product of the resistivity $\rho$ and the mean free path $\ell$ is a material constant: $\rho\ell$ = const. The mean free path is determined by elastic scattering. At high (room) temperatures the scattering is dominated by phonons, while at low temperatures ($\leq$ 10 K) the size of the smallest characteristic imperfection (sample or/and grain boundary) defines the mean free path. For macroscopic non-single-crystalline objects typically this scale is set by the grain size. For example, in our aluminium nanowires the average crystalline grain $d_{grain}$ size is about 30 nm [12]. When the smallest dimension of a wire (e.g. height) starts to be smaller than the grain size the analysis is complicated by the size dependence of the mean free path $\ell(\sigma)$. Assuming $\ell$ =const ~ $d_{grain}$ for 'thick' wires ($min$(w, h) >> $d_{grain}$) and $\ell(\sigma)$ ~ $\sigma^{1/2}$ for the narrow ones, and utilising the tabulated value for a dirty-limit aluminium $\rho\ell$ = 6 x $10^{-16}$ $\Omega m^2$ (e.g. [13]) we were able to find a reasonable agreement between the SPM-measured and the resistance-calculated cross-section $\sigma$. Taking into consideration the concrete evolution of the shape of a nanostructure (e.g. flattening scenario) and the Ohm's law $R=\rho L/\sigma \equiv [\rho\ell(\sigma)] L / [\ell(\sigma) \sigma]$ one can estimate the 'expected' value of the resistance of a nanowire subjected to a certain ion fluence (figure 6). The described approach gives a reasonable estimation of a wire shape, dimension and resistance evolution while sputtering. The resistance of the studied samples was always measured in a dedicated (cryogenic) set-up separate from the sputtering chamber. For correct measurement of electric resistance of a sub-100 nm nanostructure it is mandatory to carefully shield and to filer all electric lines from the 'noisy' electromagnetic environment. Three stages of RLC filters were used, where the last one (double-T RC filter) was integrated in the sample holder operating from room to cryogenic temperatures. Additional complicity comes from charging of a sample while the ion beam treatment. Neither the bonding wires, nor the contact pads on *Si/SiO$_x$* chip were ever destroyed by the ion beam sputtering. While the thinnest parts of the nanostructures were often 'burnt' if no precautions were taken. We do not have enough statistics to associate these damages exclusively to charge accumulation while sputtering. Nanostructures are known to be very sensitive to static electricity. To prevent an accidental discharge, all contacts of the samples were always grounded except the periods of measuring sessions. The shortcuts were removed only after installation of the processed structures (once fixed on the sample holder) in the measuring set-up. We found these strict requirements incompatible with the general purpose vacuum chamber used for sputtering, particularly – sliding contacts of the rotating sample holder. However, we believe that for some applications resistance of a micro- or a nanostructure can be measured *in situ* just in the sputtering chamber providing a powerful tool for monitoring the sample processing. There are no principal objections against integration of a measuring circuit, sputtering system and analysing instrumentation (e.g. SEM or SPM) in a single set-up.

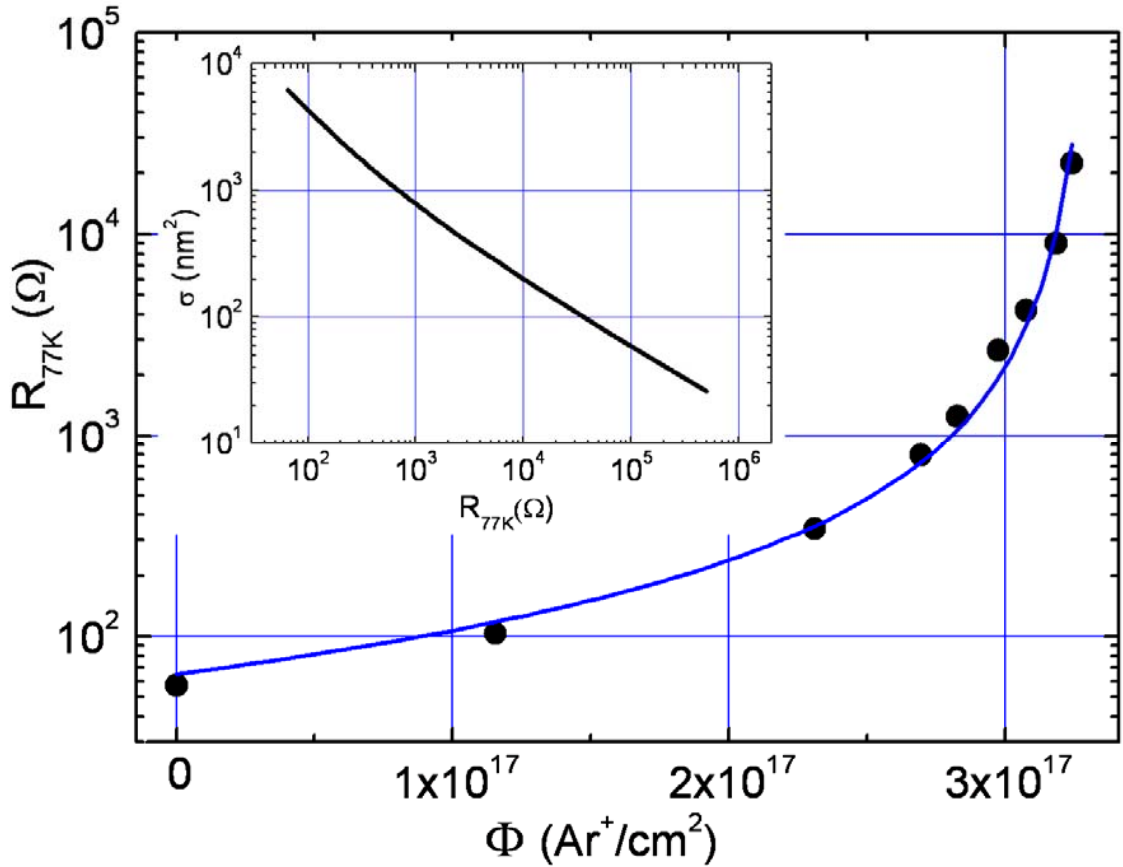

Figure 6. Resistance R of aluminium nanowire at 77 K vs. sputtering dose $\Phi$. The original dimensions of the L=10 μm long sample are: $\sigma$ = 6200 nm$^2$, height h= 58 nm, side wall angle $\theta$=65°. For details of the geometry see figure 4. Dots are experimental data. Line represents the calculation with two fitting parameter $\rho\ell$ and $V_{top}$. For numerous aluminium samples the best fit value has been found to be $\rho\ell$= (6±1) x 10$^{-16}$ $\Omega$m$^2$. The fitted sputtering rate $V_{top}$ is in a good agreement with calibration from a 2D aluminium film. Inset shows the same data where the experimentally measured resistance is re-calculated into the wire cross-section.

## 4. Conclusion

We have shown that low energy ion beam sputtering can be used for down-sizing of pre-fabricated nanostructures. The method is complementary to existing nanofabrication techniques providing cheap and reliable post-processing enabling sub-10 nm feature formation. The described technique can be utilized for a wide variety of materials whenever there is a need to reduce the size or/and reshape nanostructures in controllable and homogenous way. The method was primarily developed for academic research where the development of quantum size phenomenon has been studied down to sub-10 nm dimensions [5-9]. The approach eliminates uncertainty typical for samples fabricated in separate runs: if there were no structural imperfections in the original sample, they cannot be introduced later by the low energy ion bombardment. On the contrary, the inevitable surface roughness is effectively reduced (figures 1 and 2). The authors believe that the method can be used also for industrial applications. Naturally, the level of integration of micro- or nanocomponents cannot be increased by the downsizing. However, in particular applications, where the extreme small dimensions or/and high aspect ratio are an issue, the approach might appear to be useful. High accuracy of the sputtering rate (as low as 1 nm/min) and compatibility of the process with high vacuum and clean room requirements makes it a powerful tool for future development of various nanoelectronic applications.

**Acknowledgment**

The authors would like to acknowledge Dr. P. Prus for valuable discussions and financial support of the EU Commission FP6 NMP-3 project 505457-1 ULTRA-1D "Experimental and theoretical investigation of electron transport in ultranarrow 1-dimensional nanostructures".